\documentclass[pra,twocolumn]{revtex4}

\usepackage{graphicx}
\usepackage{hyperref}
\usepackage{amssymb,amsmath}
\usepackage{array}
\usepackage{color}

\begin{document}

\title{Type-II Optical Parametric Oscillator :\\
a versatile source of quantum correlations and entanglement}

\author{Julien Laurat, Thomas Coudreau and Claude Fabre{\footnote[1]{To whom
correspondence should be addressed (fabre@spectro.jussieu.fr)}}}

\affiliation{Laboratoire Kastler Brossel, Universit{\'e} P. et M.
Curie, Case 74, 4 Place Jussieu, 75252 Paris cedex 05, France}

\begin{abstract}
Type-II Optical Parametric Oscillators are efficient sources of
quadrature squeezed or polarization-squeezed light, intensity
correlated beams, and entangled light. We review here the
different levels of quantum correlations and entanglement that are
reached in this device, and present some applications.
\end{abstract}

\date{\today}

\maketitle

\section{Introduction}

Quantum correlations play a key role in quantum mechanics, in
basic issues such as non-locality or decoherence and also in
potential applications such as quantum information processing and
computation. The existence of correlations between different
physical systems is obviously not a specific property of quantum
physics : it is simply the consequence of a former interaction,
whatever its character, between the systems submitted to the
measurement. Consequently, the observation or prediction of a
correlation, even perfect, between the measurements of two
variables is not at all a proof of the quantum character of the
phenomenon under study. One can find in the literature a great
deal of criteria setting a border between the classical and the
quantum effects, differing by the definitions of the quantum
character of a given physical situation. The purpose of this paper
is to review some criteria for quantum correlations (section
\ref{criteres}) and to describe how a single device, namely a
type-II Optical Parametric Oscillator (section \ref{montage})
produces various kinds of correlations fulfilling these criteria
(sections \ref{2quad} and \ref{4quad}). We will also show that the
same device can generate entangled states in a non-standard form
(section \ref{manipulation}). This last section will provide a
good insight into general properties of two-mode gaussian states,
illustrated in terms of covariance matrices.

The results presented in this paper are detailed in Refs.
\cite{criteres,optlett_above,mescond,theo_mescond,pra_bf,pra_epr}.

\section{Correlation criteria}\label{criteres}

Let us consider two light beams denoted by indices 1 and 2. We
denote by $\delta X_{1,2}$ one quadrature component of these
beams, which can be measured either by direct photodetection
(amplitude quadrature) or by an homodyne detection, normalized in
such a way that vacuum fluctuations have a variance equal to 1. We
restrict ourselves in this paper to the "balanced" case when the
two beams have equal variances $F$ on these quadratures, and also
equal frequencies. More general criteria in the unbalanced case
can be found in \cite{criteres}. Let us stress also that we are in
the case where the quantum properties of the system are well
described by a linearized approach of quantum fluctuations.

\subsection{"Gemellity"}

A first criterion of quantum correlations can be defined as
follows: \textit{the correlation measured in the system cannot be
described by a semi-classical model involving classical
electromagnetic fields having classical fluctuations}.

It is easy to show that the classical character of light fields is
preserved by linear "passive" optical devices, which involve only
linear, energy-preserving, optical elements like beamsplitters and
free propagation. In order to ascertain the quantum character of
correlations existing between $\delta X_{1}$ and $\delta X_{2}$,
the simplest way is therefore to process the two beams by all
possible linear passive optical devices : if one is able to
produce in such a way a beam having fluctuations below the quantum
noise limit, that is well-known to be "non-classical", the initial
correlation will also be termed as "non-classical".

For balanced beams, the best linear processing is simply to send
them on a 50/50 beam-splitter: one obtains at one of its output
ports a beam with quadrature fluctuations $\delta X_{out}$ given
by
\begin{equation}
\delta X_{out} = \frac{\delta X_1 - \delta X_2}{\sqrt 2}
\end{equation}
having a variance given by:
\begin{equation}
G = \frac{1}{2}\left\langle \left(\delta X_1 - \delta X_2\right)^2
\right\rangle
\end{equation}
The correlation will be said to be non-classical when this
quantity, that can be called the "gemellity", is smaller than 1.
$G$ can also be written in terms of the  noise variance of each
beam $F$ and of the normalized correlation coefficient $C_{12}$:
\begin{equation}
G = F (1 - \left|C_{12}\right|).\label{eq:gemellite}
\end{equation}
Therefore a correlation is non-classical when the normalized
correlation function fulfills the following condition:
\begin{equation}
|C_{12}| > 1 - \frac{1}{F}
\end{equation}
Thus the larger the classical noise is on each beam, the more
stringent the condition becomes.

Finally, let us stress that $G$ can be easily measured
experimentally: this is done in all homodyne detection schemes of
squeezing, which actually measures the quantum character of the
correlation existing between the two beams produced by mixing the
field to measure with the local oscillator, and in all twin beams
experiments\cite{optlett_above,traditional}.

\subsection{Quantum Non Demolition correlation}

When two observables $M_1$ and $M_2$ are correlated, the
measurement of $M_2$ gives some information about the value of
$M_1$ without any interaction with system $1$. Correlations
provide therefore opportunities for Non Demolition measurements.
One is led to a second criterion of quantum correlation:
\textit{the correlation is such that the information extracted
from the measurement on one field provides a Quantum Non
Demolition measurement of the other \cite{qnd}}.

This criterion is related to the conditional variance given by~:
\begin{equation}
V_{1|2} = F_1(1-C_{12}^2).\label{eq:varcond}
\end{equation}
where $F_1$ is the noise of beam 1 normalized to shot noise. QND
correlations correspond to values of $V_{1|2}$ below 1, and
therefore to a correlation satisfying the inequality:
\begin{equation}
|C_{12}| > \sqrt{1-\frac{1}{F}}
\end{equation}

Eq. \ref{eq:varcond} can also be expressed in terms of the
gemellity:
\begin{equation}
V_{1|2} = V_{2|1} =
G(1+|C_{12}|)=2G-\frac{G^2}{F}.\label{eq:varcond2}
\end{equation}
It is easy to show from these relations that all QND-correlated
beams have a gemellity smaller than 1, whereas a gemellity smaller
than 0.5 is required to have QND-correlated beams (in the limit of
large individual noise).

\subsection{Inseparability}
\label{sec:insep}

Let us now define a new criterion related to entanglement:
\textit{the correlation cannot be described by separable quantum
states}. Can the state be written as (a sum of) tensor products or
not?

If one is sure that the system is in a pure state, separable or
factorizable state vectors give rise to no correlations at all,
whatever the observables: the existence of a non-zero correlation,
even "classical", on a single quadrature is sufficient to prove
the inseparability of the state.

When the state is mixed, which is the general case, this is no
longer the case. Let us consider for example the mixed state
described by the density matrix
\begin{equation}
\rho = \sum_n p_n \left(|1: n\rangle \otimes |2: n \rangle\right)
\left(\langle 1:n| \otimes \langle 2:n |\right)
\end{equation}
where $|1,2: n \rangle$ is a Fock state with $n$ photons in mode
1, 2. This highly non-classical state has perfect intensity
correlations ($C_{12}=1$), so that $G=V_{1|2} = 0$). However, it
is a separable state, being a statistical mixture of factorized
state vectors.

In order to ascertain the separable character of the physical
state of a system, one needs to make two joint correlation
measurements on non-commuting observables on the system, and not
only one, as was the case in the two previous criteria. More
precisely, Duan \emph{et al.}\cite{duan} have shown that, in the
case of Gaussian states for which the covariance matrix is
expressed in the so-called standard form, there exists a necessary
and sufficient criterion of separability in terms of the quantity
$\mathcal I$, that we will call "separability", and is given by :
\begin{eqnarray}
\mathcal I &=& \frac{1}{4} \left(\left\langle \left(\delta X_1 -
\delta X_2\right)^2\right\rangle + \left\langle \left(\delta P_1 +
\delta
P_2\right)^2\right\rangle\right)\nonumber\\&=&\frac{1}{2}\left(G_X+G_P\right).
\label{eq:insep}
\end{eqnarray}
The separability appears as the half-sum of the gemellity
measuring the correlations between quadratures $\delta X$ and the
(anti)gemellity measuring the anticorrelations between $\delta P$.
A state for which $\mathcal I$ is smaller than one will be a
non-separable or entangled state. As a consequence, classically
correlated beams, for which these two gemellities are larger than
1, are separable.

Let us note that in the case of symmetric gaussian states the
entanglement can be quantified by a quantity called entropy of
formation -- or entanglement of formation $EOF$ --, that was
introduced in Ref.\cite{giedke}. It represents the amount of pure
state entanglement needed to prepare the entangled state. This
quantity is related to the value of the inseparability $\mathcal
I$ by:
\begin{equation}\label{EOF1}
EOF=c_{+}\log_{2}(c_{+})-c_{-}\log_{2}(c_{-})
\end{equation}
with
\begin{equation}\label{EOF2}
c_{\pm}=(\mathcal{I}^{-1/2} \pm \mathcal{I}^{1/2})^{2}/4
\end{equation}
EOF takes a positive value only for entangled beams. Its interest
is that it constitutes a real measure of the amount of
entanglement. In addition, it is also used in the discrete
variable regime.

A more general quantity has been introduced to characterize the
entanglement: the logarithmic negativity \cite{vidal}. This
quantity can be calculated for any arbitrary bipartite system. We
will consider it in more detail in the last section of this paper
where the generated two-mode state is not in a standard form.

\subsection{Einstein-Podolsky-Rosen correlations}

Two correlations give the opportunity of Non Demolition
measurements on two non-commuting variables. As for a single
quadrature, one can be interested in the quality of the
information that one gets on one beam by measuring the other. This
question is related to the question asked by Einstein, Podolsky
and Rosen in their famous paper \cite{epr35}. In particular, we
will say that we have \textit{"EPR beams" when the information
extracted from the the measurement of the two quadratures of one
field provide values for the quadratures of the other which
"violate" the Heisenberg inequality.} This criterion was
considered and discussed extensively by Reid and
co-workers\cite{reid}. They showed that to characterize this
property, one can use the product of the conditional variances,
\begin{equation}
\mathcal V = V_{X_1|X_2} . V_{P_1|P_2}.
\end{equation}
When this quantity is smaller than one we will say that we have
"EPR-correlated beams". Let us note that when this condition is
fulfilled, one can perform double QND-measurements, that is two
QND-measurements on non-commuting quadratures. One can show that
all EPR-correlated beams are not separable, whereas the reverse is
not true. EPR-correlation is therefore the strongest of the
correlation criteria that we have listed here. One can envision
other criteria which are even stronger, but not relevant for the
problem of measuring correlated quadratures with Gaussian
statistics that we are considering here.

\section{Experimental investigation of quantum correlations}
\label{experiences}

In this section, we will show how these various criteria can be
tested using the states produced by a triply resonant type-II
Optical Parametric Oscillator. Such a system consists of a
triply-resonant optical cavity containing a type-II phase matched
$\chi^{(2)}$ crystal. Spontaneous parametric down-conversion which
occurs in such crystals is well known to produce twin photons,
that is photons created in pairs. When such a crystal is placed
inside a cavity and the system pumped above a critical value
(threshold), one generates intense beams which are correlated. The
system transfers the correlations existing in the discrete regime
to the continuous variable one. The OPO is thus an ideal system to
test the various criteria that we have mentioned above.

\begin{figure}[h]
\centerline{\includegraphics[width=0.9\columnwidth]{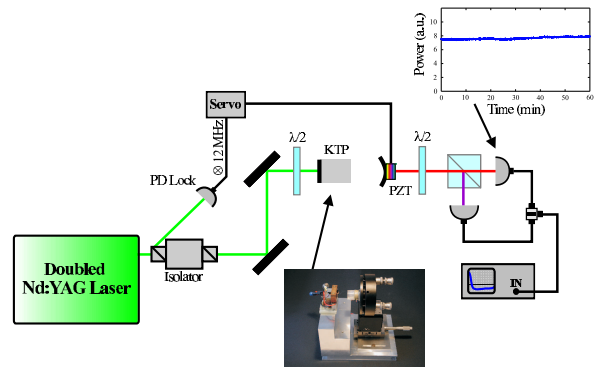}}
\caption{A cw doubled Nd:YAG laser pumps above threshold a type-II
OPO. Intensity correlations are directly measured by a balanced
detection scheme. PD Lock: FND-100 photodiode for locking of the
OPO.}\label{setup_twin}
\end{figure}

\subsection{Experimental set-up}
\label{montage}

The experimental setup is shown in Fig. \ref{setup_twin}. A
continuous-wave frequency-doubled Nd:YAG laser  pumps a triply
resonant OPO above threshold, made of a semi-monolithic linear
cavity. The intensity reflection coefficients for the input
coupler are 95\% for the pump at 532 nm and almost 100\% for the
signal and idler beams at 1064 nm. The output mirror is highly
reflective for the pump and its transmission coefficient $T$ can
be chosen to be 5 or 10\%. With $T=5\%$, at exact triple
resonance, the oscillation threshold is less than 15 mW. In spite
of the triple resonance which generally makes OPOs much more
sensitive to disturbances, length and temperature controls enable
stable operation over more than one hour without mode-hopping.

\subsection{"$2 \times 1$ quadrature" case}
\label{2quad}

\subsubsection{Twin beams}\label{sec:twin}

Type II optical parametric oscillators are well-known to generate
above threshold highly quantum correlated bright twin beams.
Intensity correlations were experimentally observed several years
ago and applied to measurements of weak physical
effects\cite{traditional}. We describe here a recent improvement
of the observed correlation.

\begin{figure}[h]
\centerline{\includegraphics[width=0.85\columnwidth]{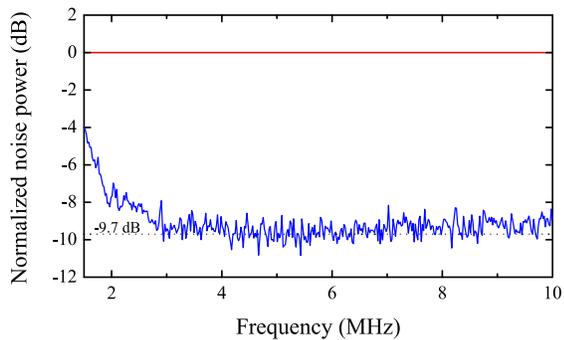}}
\caption{Normalized noise power of the intensity difference of the
signal and idler as a function of the frequency, after correction
of the electronic noise.}\label{result_twin}
\end{figure}

Intensity correlations are directly measured by a balanced
detection scheme (Fig. \ref{setup_twin}). The signal and idler
orthogonally polarized beams are separated on a polarizing beam
splitter and detected on a pair of high quantum efficiency
photodiodes. With a transmission $T=10\%$ for the output mirror,
we have obtained a noise reduction of $9.7 \pm 0.5$ dB (89\%)
around 5 MHz (Fig. \ref{result_twin}), which corresponds to a
gemellity of $G=0.11$. To the best of our knowledge, this noise
reduction is the strongest reported to date in the experimental
quantum optics field.

\subsubsection{QND correlations and conditional preparation of a non-classical state}

The observed correlation is strong enough to yield a conditional
variance well below 1. We will show now that such a
QND-correlation can be used to produce a squeezed state \emph{via}
conditional preparation performed on continuous variables.

A well-known technique to generate a single photon state from twin
photons is to use the method of conditional measurement: if one
labels (1) and (2) the two modes in which the twin photons are
emitted, it consists in retaining in the information collected in
mode (1) only the counts occurring when a photon is detected in
mode (2) within a given time window $\Delta T$. State preparation
by conditional measurement can be readily extended to the
continuous variable regime, where the instantaneous values of the
signal and idler photocurrents play the role of the occurrence of
counts in the photon counting regime. The technique consists in
selecting the signal photocurrent $I_s$ only during the time
intervals when the idler intensity $I_i$ has a given value $I_0$
(within a band $\Delta I$ smaller than the photocurrent standard
deviation). The measurements outside these time intervals are
discarded. If the correlation is perfect and the interval $\Delta
I$ close to zero, the recorded signal intensity is perfectly
constant, and an intense number state is generated; in a real
experiment, the correlation between the signal and idler
photocurrents is not perfect, and the selection band $\Delta I$ is
finite, so that the method will not prepare a perfect number
state, but a sub-Poissonian state instead.

A theoretical analysis of this protocol \cite{theo_mescond} shows
that in the limit where $\Delta I$ is very small the conditional
measurement produces a beam characterized by a Fano factor equal
to the conditional variance of the signal and idler beams. This
means that the present protocol produces a sub-Poissonian beam
when the signal and idler beams are QND-correlated. As shown in
Eq. \ref{eq:varcond2} in the limit of large correlations, the
residual intensity noise $F$ on the conditionally prepared state
will be equal to twice the gemellity ($F=V \simeq 2 G$).

\begin{figure}
\centerline{\includegraphics[width=.85\columnwidth]{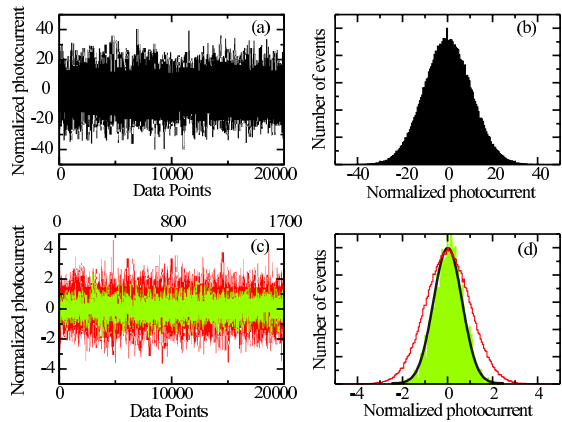}}
\caption{\label{courbes_exp}Experimental results: (a) Idler
intensity fluctuations: 200 000 acquired points at 3.5 MHz
demodulation frequency (only 20 000 shown). (b) Corresponding
probability distribution. The unit is the width $\sigma_0$ of the
Poisson distribution of same mean intensity (shot noise). (c)
Values of the signal intensity conditionally selected by the value
of the idler intensity recorded at the same time (selection
bandwidth $\Delta I$ equal to 0.1 $\sigma_0$ around the mean),
superimposed to the corresponding experimentally measured shot
noise. (d) Corresponding probability distribution, compared to the
Poisson distribution (grey line), displaying the sub-Poissonian
character of the conditionally generated state. The black line is
a gaussian fit of the intensity distribution.}
\end{figure}

Figure \ref{courbes_exp} sums up the experimental results. The
Fano factor $F$  of both the signal and idler beams exceeds 100
(20~dB above the shot noise level), and the measured gemellity $G$
is equal to $0.18$ ($0.14$ after correction of dark noise). The
ensemble of values of the signal intensity for which the idler
intensity falls within the selection band is given in figure
\ref{courbes_exp} (c): one indeed observes a significant narrowing
of the probability distribution below the shot noise level. With a
selection bandwidth $\Delta I$ equal to 0.1 times the standard
deviation $\sigma_0$ of a coherent state having the same power
(shot noise level), the conditionally prepared light state has a
measured Fano factor $F=0.36$, which turns out to be equal, as
expected, to the conditional variance of the twin beams. The
success rate of the conditional preparation is around 0.85\% (1700
points out of 200 000 are accepted). An advantage of the
conditional preparation using continuous variables is that one can
use at the same time different selection non-overlapping bands on
the idler beam. Each allows one to conditionally prepare a
different sub-Poissonian state, each having a Fano factor
$F=0.36$. With 200 different selection bands, the overall success
rate is close to $100\%$.

\subsection{"$2\times 2$" quadratures case} \label{4quad}

\subsubsection{Entanglement below threshold}

\begin{figure*}
\centerline{\includegraphics[width=1.5\columnwidth]{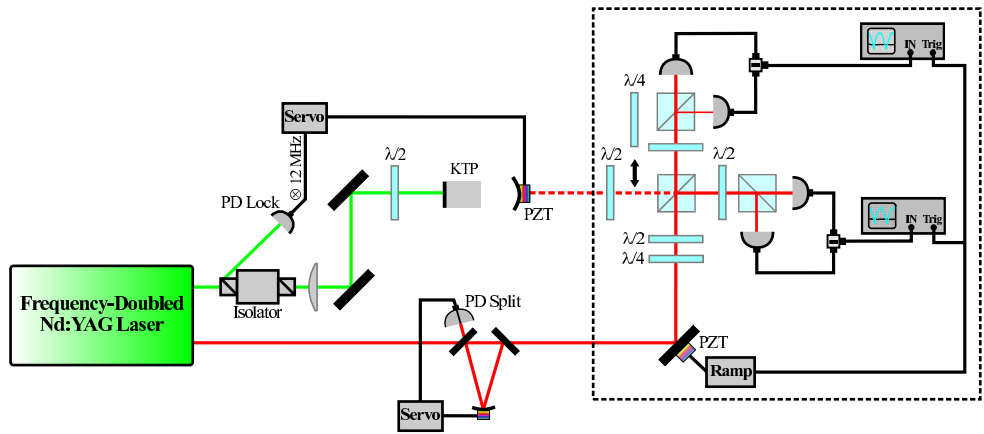}}
\caption{A doubled Nd:YAG laser pumps a type II OPO, below or
above threshold. The generated two-mode state is characterized by
two simultaneous homodyne detections. PD Split: split two-element
InGaAs photodiode for tilt-locking of the filtering
cavity.}\label{setup}
\end{figure*}

Type-II OPO below threshold are well-known to generate entangled
beams. The first experimental demonstration of EPR correlations in
the continuous variable regime in 1992 was performed with such a
device\cite{ou92}. Our experimental setup is similar to the
previous one (Fig. \ref{setup_twin}) but the detection system is
now based on two simultaneous homodyne detections (Fig.
\ref{setup}). In order to measure the separability $\mathcal{I}$,
one must characterize the noise of the superposition modes
oriented $\pm45^{\circ}$ from the axes of the crystal:
\begin{eqnarray} A_{+}=\frac{A_{1}+A_{2}}{\sqrt{2}} \qquad
\textrm{and} \qquad A_{-}=\frac{A_{1}-A_{2}}{\sqrt{2}} \nonumber
\end{eqnarray}
Eq. \ref{eq:insep} shows that the signal and idler fields are
entangled as soon as these two modes have squeezed fluctuations on
orthogonal quadratures. The orthogonally polarized modes are
separated on a first polarizing beam splitter at the output of the
OPO. A half-wave plate inserted before this polarizing beam
splitter enables us to choose the fields to characterize: the
signal and idler modes which are entangled, or the $\pm45^{\circ}$
rotated modes which are squeezed. The detection setup is able to
characterize simultaneously the two chosen modes with the same
phase reference, and to measure the noise reductions either in
quadrature ("in phase homodyne detection") or in phase ("in
quadrature homodyne detection"), by inserting or not a $\lambda/4$
plate in the beam exiting the OPO. This configuration permits a
direct and instantaneous verification of the inseparability
criterion by simply adding the two squeezed variances.

\begin{figure}[htpb!]
\centerline{\includegraphics[width=.95\columnwidth,clip=]{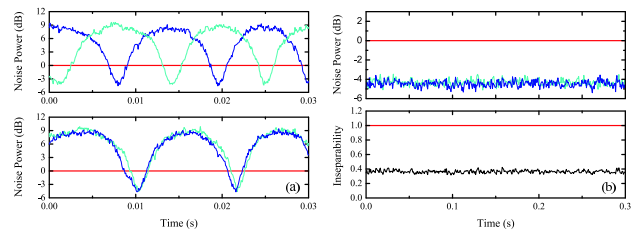}}
\caption{(a) Normalized noise variances at 3.5 MHz of the $\pm
45^{\circ}$ modes while scanning the local oscillator phase. The
first plot corresponds to in-phase homodyne detections and the
second one in-quadrature. Squeezing is well observed on orthogonal
quadratures. (RBW 100 kHz, VBW 1 kHz) (b) Normalized noise
variances at 3.5 MHz of the $\pm 45^{\circ}$ modes and
inseparability $\mathcal I$ for signal and idler modes. The
homodyne detections are in-quadrature and locked on the squeezed
quadratures. After correction of the electronic noise, the
inseparability criterion reaches $0.33\pm 0.02$. (RBW 100 kHz, VBW
300 Hz).}\label{verysmallscan}
\end{figure}

Typical spectrum analyzer traces while scanning the local
oscillator phase are shown on Fig. \ref{verysmallscan}. Normalized
noise variances of the $\pm 45^{\circ}$ vacuum modes at a given
noise frequency of 3.5 MHz are superimposed for in-phase and
in-quadrature homodyne detections. One indeed observes, as
expected, correlations and anti-correlations of the emitted modes
on orthogonal quadratures. The homodyne detection can be locked on
the squeezed quadrature (Figure \ref{verysmallscan}). The observed
amount of simultaneous squeezing for the two rotated modes is
$-4.3 \pm 0.3$ dB and $-4.5 \pm 0.3$ dB below the standard quantum
limit ($-4.7 \pm 0.3$ dB and $-4.9 \pm 0.3$ dB after correction of
the electronic noise). This gives a value of the separability of
$\mathcal I=0.33 \pm 0.02$, well below the unit limit for
inseparability. With a measured value of the parameter $F$ of
$6.6$, one obtains a product of conditional variances of $0.42\pm
0.05$, well below 1, which establishes the EPR character of the
measured correlations. The entanglement of formation $EOF$ of the
two beams is, according to formulae (\ref{EOF1}) and (\ref{EOF2}),
equal to $1.1 \pm 0.1\,ebits$. To the best of our knowledge, this
setup generates the best EPR/entangled beams to date produced in
the continuous variable regime.

\begin{figure}[htpb!]
\centerline{\includegraphics[width=0.85\columnwidth,clip=]{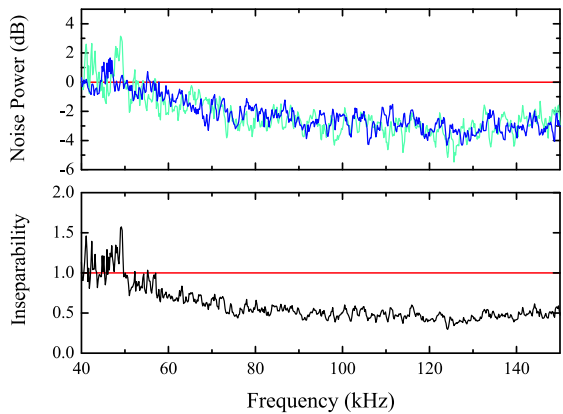}}
\caption{\label{bf}Normalized noise variances from 40 kHz to 150
kHz of the $\pm 45^{\circ}$ modes after correction of the
electronic noise and inseparability criterion for signal and idler
modes. Squeezing and entanglement are observed down to 50 kHz.
(RBW 3 kHz, VBW 10 Hz)}
\end{figure}

Non-classical properties are generally measured in the MHz range
of Fourier frequencies, because of the presence of large classical
noise at lower frequencies. In the present device significant
quantum correlations and EPR entanglement have been observed in
from 50 kHz to 10 MHz. Figure \ref{bf} gives the squeezed
variances for low noise frequencies, between 40 kHz and 150 kHz.
Let us mention that squeezing from a single type-I OPA was
recently reported at a record Fourier frequency of 200 Hz
 \cite{200Hz}.

\subsubsection{Bright EPR beams above threshold and polarization
squeezing}
\label{phlock}

A type-II OPO pumped above threshold has been theoretically
predicted to be a very efficient source of bright entangled and
EPR beams. This means that, in addition to the already
demonstrated intensity correlations, phase anticorrelations exist
in the system. However, they can be easily measured by usual
homodyne detection techniques only in the frequency-degenerate
regime. Frequency degeneracy occurs only accidentally above
threshold because it corresponds to a single point in the
experimental parameter space. Actually, up to now, no direct
evidence of such phase anti-correlations has been observed. In
1998, Mason and Wong proposed an elegant way to achieve frequency
degenerate operation above threshold \cite{Mason98,Fabre99}: they
inserted inside the OPO cavity a birefringent plate making an
angle with the axis of the non-linear crystal. The induced linear
coupling between the signal and idler results in a locking
phenomenon\cite{Pikovsky}. It has been shown theoretically that in
such a "self-phase-locked" OPO the quantum correlations are
preserved for small angles of the plate and that the system
produces entangled states in a wide range of parameters
\cite{longcham1,longcham2}.

\begin{figure}
\centerline{\includegraphics[width=0.85\columnwidth]{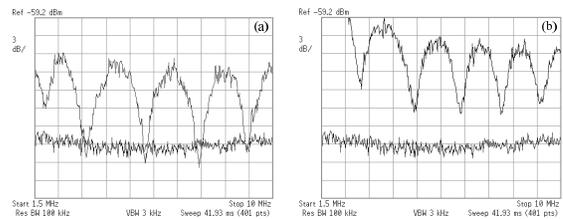}}
\caption{(a) Noise power of the mode $A_-$ while scanning
simultaneously the phase of the local oscillator and the noise
frequency between 1.5 and 10 MHz. The lower trace gives the shot
noise level. (b) Noise power of the mode $A_+$ while scanning the
phase of the local oscillator, for a noise frequency between 1.5
and 10 MHz. The shot noise level is given by the lower trace plus
3 dB.}\label{audessus}
\end{figure}

In the experiment, the frequency locking phenomenon can be
maintained during more than hour. Degenerate operation is
confirmed by the fact that the generated mode has now a fixed
polarization: at the minimum threshold point, the generated state
is linearly polarized at +45$^{\circ}$. Due to the defined phase
relation existing now between the signal and idler fields, $A_+$
is a bright mode, and $A_-$ has a zero mean value. Figure
\ref{audessus} (a) gives the noise power of the mode $A_-$ while
scanning the local oscillator phase, for a transmission $T=5\%$
and a plate angle of 0.1$^{\circ}$. A noise reduction of 4.5dB is
observed. This strong noise reduction on the mode $A_-$ confirms
the quantum intensity correlation between the signal and idler
modes. Figure \ref{audessus} (b) shows the noise power of the mode
$A_+$ in the same condition. As the plate angle is very small, a
similar amount of noise reduction is expected. However, a slight
excess noise of 3 dB is measured for the minimal noise quadrature:
the phase anticorrelations appear to be slightly degraded,
probably by external noise sources.

Despite this slight excess noise which prevents from reaching the
proof of entanglement in the OPO above threshold, the generated
state turns out to be squeezed in the polarization orthogonal to
the mean field: $A_+$ is the main mode and $A_-$ the squeezed
vacuum one. This condition is required to obtain a so-called
"polarization squeezed" state \cite{Korolkova,Bowen,Josse}. 4.5 dB
of polarization squeezing has been thus generated in the
self-phase-locked OPO. Such states have recently raised great
interest, in particular because of the possibility to map quantum
polarization state of light onto an atomic ensemble \cite{Hald}.

\begin{figure}[htpb!]
\centerline{\includegraphics[width=.8\columnwidth]{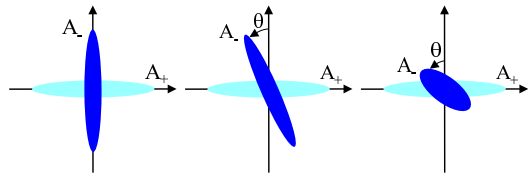}}
\caption{\label{ellipse} Fresnel representation of the noise
ellipse of the $\pm 45^{\circ}$ rotated modes when the plate angle
is increased. Without coupling, squeezing is predicted on
orthogonal quadratures. The noise ellipse of the $-45^{\circ}$
mode rotates and the noise reduction is degraded when the coupling
increases while the $+45^{\circ}$ rotated mode is not affected.}
\end{figure}

\begin{figure*}[t]
\small
\begin{eqnarray}
\Gamma_{A_{1}\,\!A_{2}} = \left( \begin{array}{cc|cc}
181.192 & 0 & 179.808 & -0.255 \\
0 & 0.386 & -0.255 & -0.383 \\
\hline 179.808 & -0.255 & 181.192 & 0\\
-0.255 & -0.383 & 0 & 0.386
\end{array} \right)
&\Longrightarrow& \Gamma'_{A_{1}\,\!A_{2}} = \left(
\begin{array}{cc|cc}
180.839 & 0 & 180.161 & 0 \\
0 & 0.739 & 0 & -0.736 \\
\hline 180.161 & 0 & 180.839 & 0\\
0 & -0.736 & 0 & 0.739
\end{array} \right)
\nonumber
\end{eqnarray}
\begin{eqnarray}
\Gamma_{A_{+}\,\!A_{-}} = \left( \begin{array}{cc|cc}
361 & 0 & 0 & 0 \\
0 & 0.00277 & 0 & 0 \\
\hline 0& 0 & 1.383 & -0.256\\
0 & 0 & -0.256 & 0.770
\end{array} \right)
&\Longrightarrow& \Gamma'_{A_{+}\,\!A_{-}} = \left(
\begin{array}{cc|cc}
361 & 0 & 0 & 0 \\
0 & 0.00277 & 0 & 0 \\
\hline 0& 0 & 0.677 & 0\\
0 & 0 & 0 & 1.476
\end{array} \right)\nonumber
\end{eqnarray}
\caption{Numerical example of covariance matrix of the
$A_{1}$/$A_{2}$ modes and the $A_{+}$/$A_{-}$ modes before and
after the non-local operation for a plate angle of
$\rho=1.3^{\circ}$. ($\sigma=0.9$ and $\Omega=0$)} \label{matrix}
\end{figure*}

\section{Manipulating entanglement with polarization elements}
\label{manipulation}

The self-phase-locked OPO can also be operated below threshold. It
produces a two-mode state with strong quantum features which
manifest themselves in terms of noise reduction properties in a
given polarization basis, and in terms of entanglement and EPR
correlations in another. This last section is devoted to the
general study of this two-mode Gaussian quantum state.

\subsection{Manipulation of entanglement in the two-mode state produced by the type-II OPO with mode coupling}

In a standard OPO the correlated quadratures are orthogonal to the
anti-correlated ones which results in squeezing of the rotated
modes on orthogonal quadratures. It is no more the case when a
linear coupling is introduced. When the plate angle increases, the
correlated quadratures rotate and the correlations are degraded.
The evolution is depicted in Fig. \ref{ellipse} through the noise
ellipses of the rotated (squeezed) modes. In order to maximize the
entanglement between the signal and idler modes, the optimal
quadratures have to be made orthogonal \cite{Wolf}. Such an
operation consists of a phase-shift of $A_{-}$ relative to
$A_{+}$. This transformation is passive and "non-local" in the
sense of the EPR argument: it acts simultaneously on the two
considered sub-systems. In the type II OPO, such "non-local"
transformations are easy to perform by inserting polarizing
birefringent elements in the total beam, because the two
polarization modes are produced by the OPO in the same transverse
spatial mode.

As the generated two-mode state is not in the standard form, we
need to use a general measure of entanglement. Let us introduce
the covariance matrix formalism and the logarithmic negativity. In
a given mode basis, the quantum properties of the generated state,
of zero mean value, are completely contained in the covariance
matrix $\Gamma_{A\,\!B}$ defined as:
\begin{equation}
\Gamma_{A\,\!B}=\left( \begin{array}{cc}
 \gamma_{A} & \sigma_{A\,\!B} \\
 \sigma_{A\,\!B}^{T} & \gamma_{B}
\end{array} \right)\nonumber
\end{equation}
$\gamma_{A}$ and $\gamma_{B}$ are the covariance matrix of the
individual modes while $\sigma_{A\,\!B}$ describes the intermodal
correlations. The elements of the covariance matrix are written
$\Gamma_{ij}=\langle \delta R_{i}\delta R_{j}+\delta R_{j}\delta
R_{i}\rangle/2$ where
$R_{\{i,i=1,..,4\}}=\{X_{A},Y_{A},X_{B},Y_{B}\}$. $X$ and $Y$
corresponds to an arbitrary orthogonal basis of quadratures. In
order to measure the degree of entanglement of Gaussian states, a
simple computable formula of the logarithmic negativity
$E_{\mathcal{N}}$ has been obtained \cite{vidal} (see also
\cite{Adesso} for a general overview). $E_{\mathcal{N}}$ can be
easily evaluated from the largest positive symplectic eigenvalue
$\xi$ of the covariance matrix which can be obtained from
\begin{eqnarray}
\xi^{2}=\frac{1}{2}(D-\sqrt{D^{2}-4\det\Gamma_{A\,\!B}}\,)
\end{eqnarray}
with
\begin{eqnarray}
D=\det\gamma_{A}+\det\gamma_{B}-2\det\sigma_{A\,\!B}
\end{eqnarray}
The two-mode state is entangled if and only if $\xi<1$. The
logarithmic negativity can thus be expressed by
$E_{\mathcal{N}}=-\log_{2}(\xi)$. The maximal entanglement which
can be extracted from a given two-mode state by passive operations
is related to the two smallest eigenvalues of $\Gamma$,
$\lambda_1$ and $\lambda_2$, by
$E_{\mathcal{N}}^{max}=-\log_{2}(\lambda_1\lambda_2)/2$
\cite{Wolf}.

We give here a numerical example for realistic experimental values
$\rho=1.3^{\circ}, \sigma=0.9$ and $\Omega=0$, where $\rho$ stands
for the plate angle, $\sigma$ the pump power normalized to the
threshold and $\Omega$ the noise frequency. The covariance
matrices for the $A_{1}$/$A_{2}$ modes and for the $A_{+}$/$A_{-}$
modes are given in Fig. \ref{matrix} with and without the
phase-shift. The matrix of the $A_{+}$/$A_{-}$ modes are
well-suited to understand the behavior of the device. At first,
the intermodal blocks are zero, showing that these two modes are
not at all correlated and consequently are the most squeezed modes
of the system: there is no way to extract more squeezing. But one
can also note that the diagonal blocks are not diagonalized
simultaneously. This corresponds to the tilt angle $\theta$ of the
squeezed quadrature of $A_{-}$. A phase-shift of the angle
$\theta$ permits to diagonalize simultaneously the two blocks and
to obtain squeezing on orthogonal quadratures. The logarithmic
negativity $E_{\mathcal{N}}$ has increased in the transformation
from $4.06$ to $4.53$. The maximal entanglement available has been
extracted in this way as
$E_{\mathcal{N}}^{max}=-\log_{2}(\lambda_1\lambda_2)/2=4.53$.

\subsection{Experimental optimization of entanglement}

Let us now describe how to experimentally optimize the EPR
entanglement generated by the self-phase-locked OPO below
threshold.

In order to extract the maximal entanglement, one must perform an
appropriate phase-shift on the rotated modes. This is achieved by
using an association of one $\lambda/2$ and one $\lambda/4$ plates
added at the output of the OPO. The double homodyne detection we
have developed is necessary in order to be able to characterize
simultaneously the two modes with the same phase reference.

\begin{figure}[htpb!]
\centerline{\includegraphics[width=.85\columnwidth]{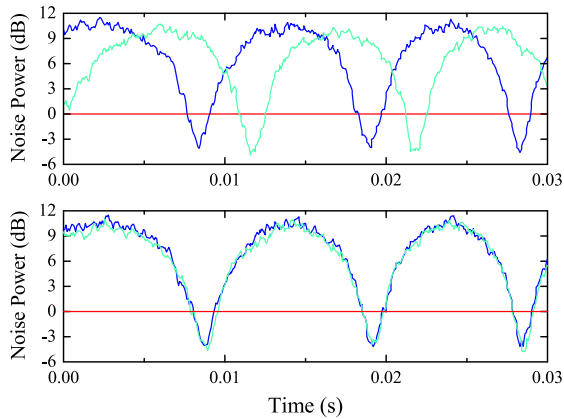}}
\caption{\label{corrige}Normalized noise variances at 3.5 MHz of
the rotated modes while scanning the local oscillator phase for a
plate angle $\rho=0.3^{\circ}$, before and after the non-local
operation. The homodyne detections are in-quadrature. After this
operation, squeezing is observed on orthogonal quadratures.}
\end{figure}

Figure \ref{corrige} displays the normalized noise variances of
the rotated modes for a plate angle of $\rho=0.3^{\circ}$, before
and after the phase-shift. The homodyne detections are operated in
quadrature so that squeezing on orthogonal quadratures is observed
simultaneously on the spectrum analyzers. After the operation is
performed, squeezing is obtained on orthogonal quadratures as in a
standard type-II OPO without mode coupling. Experimentally, the
logarithmic negativity goes from 1.13 to 1.32, showing that we are
able to extract more quantum resource from the state after the
operation.

\section{Conclusion}

We have seen that type II triply resonant OPO produce in a very
stable way the strongest intensity correlation and EPR
entanglement to date. Phase-locked, frequency degenerate operation
can be obtained using an intracavity birefringent plate. This
locking permits the experimental realization of homodyne detection
of the quadratures even when the system is operated above
threshold. This result opens a very promising way to the direct
generation of intense entangled beams and offers a new and simple
method to achieve strong polarization squeezing. Below threshold,
the self-phase-locked OPO exhibits a very rich and interesting
behavior which provides a good insight into entanglement
manipulation by passive operations. This opens the way to the
manipulation and optimization of quantum properties in highly
multimode Hilbert spaces.

\end{document}